\documentclass{PoS}

\usepackage[sumlimits,intlimits,namelimits]{amsmath}
\usepackage{mathtools}
\usepackage{graphicx}
\usepackage{grffile}
\usepackage[noabbrev]{cleveref}
\usepackage{xcolor}
\usepackage{subfig}
\usepackage{macros}
\usepackage[backend=biber,
			backref=false,
			style=numeric-comp,
			sorting=none,
			block=ragged,
			firstinits=true
			]{biblatex}

\renewbibmacro*{pageref}{}
\renewbibmacro*{in:}{\printtext{\vphantom{a}}}
\DeclareFieldFormat{pages}{#1}
\DefineBibliographyStrings{english}{mathesis = {Master's thesis}}
\providecommand{\texorpdfstring}[2]{#1}
\usepackage{todonotes}

\title{Recent Results on Light-Meson Spectroscopy from COMPASS}

\ShortTitle{Recent Results on Light-Meson Spectroscopy from COMPASS}

\author{\speaker{Stefan Wallner}\thanks{For the COMPASS collaboration}\\
        Institute for Hadronic Structure and Fundamental Symmetries (TU Munich)\\
        E-mail: \email{stefan.wallner@tum.de}}

\abstract{%
The main goal of the spectroscopy program at COMPASS is to explore the light-meson spectrum below about $2\,\text{GeV}/c^2$ in diffractive production. Our flagship channel is the decay into three charged pions: $p + \pi^-\to \pi^-\pi^-\pi^+ + p_\text{recoil}$, for which COMPASS has acquired the so far world's largest dataset of roughly $50\,\text{M}$ exclusive events using an $190\,\text{GeV}/c$  $\pi^-$ beam.
Based on this dataset, we performed an extensive partial-wave analysis. In order to extract the resonance parameters of the $\pi_J$ and $a_J$ states that appear in the $\pi^-\pi^-\pi^+$ system, we performed the so far largest resonance-model fit, using Breit-Wigner resonances and non-resonant contributions.

This method in combination with the high statistical precision of our measurement allows us to study ground and excited states.
We have found an evidence of the $a_1(1640)$ and $a_2(1700)$ in our data, which are the first excitations of the $a_1(1260)$ and $a_2(1320)$, respectively. The relative strength of the excited states with respect to the corresponding ground state is larger in the $f_2(1270)\,\pi$ decay mode compared to the $\rho(770)\,\pi$ decay mode.
We also study the spectrum of $\pi_2$ states in our data. Therefore, we simultaneously describe four $J^{PC}=2^{-+}$ waves in the resonance-model fit by using three $\pi_2$ resonances, the $\pi_2(1670)$, the $\pi_2(1880)$, and the $\pi_2(2005)$. Within the limits of our model, we can conclude that the $\pi_2(2005)$ is required to describe all four $2^{-+}$ waves properly.
}

\FullConference{XVII International Conference on Hadron Spectroscopy and Structure - Hadron2017\\
		25-29 September, 2017\\
		University of Salamanca, Salamanca, Spain}

\makeatletter
\let\@fnsymbol\@alph
\makeatother

\newcommand{\twoPlotWidth}{0.441\linewidth}
\newcommand{\threePlotWidth}{0.32\linewidth}
\begin{document}
\section{Light-meson spectroscopy at COMPASS}
COMPASS is a fixed-target multi-purpose experiment located at CERN (SPS).
Positive and negative secondary hadron beams with momenta of \SI{190}{\GeVc} or a tertiary muon beam of \SI{160}{\GeVc} can impinge various types of targets.
The final-state particles are detected by a two-stage magnetic spectrometer, which has a large acceptance over a broad kinematic range.

The main goal of the COMPASS spectroscopy program is to study the light-meson spectrum up to about \SI{2}{\GeVcc}.
At COMPASS, these light mesons are produced in diffractive scattering of the \SI{190}{\GeVc} negative hadron beam, which is mainly composed out of negative pions, is scattered off a liquid-hydrogen target. In these reactions, we can study $a_J$- and $\pi_J$-like mesons.
Due to their very short lifetime, we observe them only in the decay into quasi-stable final-state particles. Our flagship channel is the decay into three charged pions: \reaction.
COMPASS has acquired the so far world's largest dataset of this channel of about \SI{50}{M} exclusive events, which allows us to apply novel analysis methods~\cite{Adolph2015}.

\section{Analysis method}
\label{sec:method}

\subsection{Partial-wave decomposition}
\label{sec:method:pwd}
We employ the method of partial-wave analysis in a two-step approach.
In the first step, called partial-wave decomposition, data are decomposed into the different contributions from the partial waves~\cite{Adolph2015}.
Therefore, we parametrize the intensity distribution $\mathcal{I}(\tau)$ of the \threePi final state in turns of five-dimensional phase-space variables that are represented by $\tau$. Using the isobar approach,  $\mathcal{I}(\tau)$ is modeled as a coherent sum of partial-wave amplitudes, which are defined by the $3\pi$ quantum numbers and the decay path represented by $\alpha = \Wave{J}{P}{C}{M}{\varepsilon}{\zeta}{\PpiNeg}{L}$\footnote{Here, $J$ is the spin of the $3\pi$ state, $P$ and $C$ its parity and charge conjugation quantum number. The spin projection of $J$ along the beam axis is given by $M^\varepsilon$. The intermediate \twoPi resonance through which the decay proceeds is represented by $\zeta$. $L$ is the orbital angular momentum between the bachelor pion and the isobar.}:
\begin{equation*}
	\mathcal{I}(\tau; \mThreePi, \tpr) = \left|\sum\limits_{\alpha}^{\text{waves}} {\mathcal T_\alpha(\mThreePi, \tpr)}\, {\psi_\alpha(\tau;\mThreePi, \tpr)}\right|^2.
\end{equation*}
The decay amplitudes $\psi_\alpha$ can be calculated. This allows us to extract the partial-wave amplitudes $\mathcal T_\alpha$, which determine the strength and phase of each wave, from the data by an unbinned maximum-likelihood fit, which is performed in narrow bins of the three-pion mass \mThreePi.

The large size of the dataset allows us to bin our data in the squared four-momentum transfer \tpr as well.
By performing the partial-wave decomposition independently in $100$ narrow \mThreePi bins and $11$ \tpr bins in the range $0.5 < \mThreePi < \SI{2.5}{\GeVcc}$ and $0.1 < \tpr < \SI{1.0}{\GeVcsq}$, we extract simultaneously the \mThreePi and \tpr dependence of the partial-wave amplitudes from the data.
\subsection{Resonance-model fit}
\label{sec:method:rmf}
At the second step of this analysis, called resonance-model fit, we parameterize the \mThreePi dependence of the partial-wave amplitudes in order to extract the masses and widths of resonances appearing in the \threePi system. Details can be found in Refs.~\cite{msc_thesis_wallner,Adolph2018}.
We model the amplitude for the wave $\alpha$ as a coherent sum over components $k$ that we assume to contribute to this wave\footnote{We dropped here the $\sqrt{\mThreePi}$ factor, the phase space, and the production factor for simplicity (see Ref.~\cite{msc_thesis_wallner} for details).}:
\begin{equation*}
	\mathcal{T}_\alpha(\mThreePi, \tpr) = \smashoperator[r]{\sum\limits_{k_\alpha}}\mathcal{C}^{k_\alpha}_\alpha(\tpr)\cdot \mathcal{D}^{k_\alpha}(\mThreePi, \tpr; \zeta_k).
\end{equation*}
The \textit{dynamical amplitudes} $\mathcal{D}^k(\mThreePi, \tpr; \zeta_k)$ represent the resonant and non-resonant components. We use relativistic Breit-Wigner amplitudes for resonances and a phenomenological parameterization for the non-resonant terms\footnote{The \mThreePi dependence of the non-resonant term is $e^{-c \tilde{q}^2(\mThreePi)}$, where $\tilde{q}(\mThreePi)$ is an approximation for the two-body break-up momentum of the isobar pion system, which is also valid below threshold.}.
Each dynamical amplitude is multiplied by a \textit{coupling amplitude} $\mathcal{C}^{k_\alpha}_\alpha(\tpr)$, which determines the strength and phase with which each component contributes to the corresponding wave. We use independent coupling amplitudes for each \tpr bin. Thus, also in this analysis step, we impose no model for the \tpr dependence of the components.
We simultaneously describe all $11$ \tpr bins in one resonance-model fit, keeping the mass and width parameters of each resonance component the same in each \tpr bin. In this way, our \tpr-resolved analysis gives us an additional dimension of information, which helps to separate resonant from non-resonant components better.

Our two-step approach allows us to select a subset of waves to be included in the resonance-model fit. In this analysis, we select a subset of 14 out of the 88 partial waves with $\JPC = \zeroMP$, \onePP, \oneMP, \twoPP, \twoMP, and \fourPP quantum numbers. This subset accounts for about \SI{60}{\percent} of the total intensity. The 14 waves are parameterized by 11 resonances; \PpiEighteenhundret, \PaOne, \PaOneFourteenTwenty, \PaOnePr, \PpiOne, \PaTwo, \PaTwoPr, \PpiTwo, \PpiTwoPr, \PpiTwoPrPr, and \PaFour; and 14 non-resonant terms.

Many systematic effects might influence the fit result. To investigate these effects we performed more than 200 systematic studies to improve our model, to test the evidence for some resonance signals in our data, and to determine systematic uncertainties of extracted resonance parameters.

\section{Results of the resonance-model fit}
In the following subsections we present selected preliminary results for resonances with $\JPC=\twoPP$, \onePP, and \twoMP quantum numbers.
As the statistical uncertainties are at least one order of magnitude smaller than the systematic ones, we quote only systematic uncertainties.
All results of the resonance-model fit are discussed in Ref.~\cite{msc_thesis_wallner} and will be published in Ref.~\cite{Adolph2018}.
\subsection{$\mathbf{\JPC = \twoPP}$ resonances}
\begin{figure}[tbp]
	\subfloat[]{%
		\includegraphics[width=\threePlotWidth]{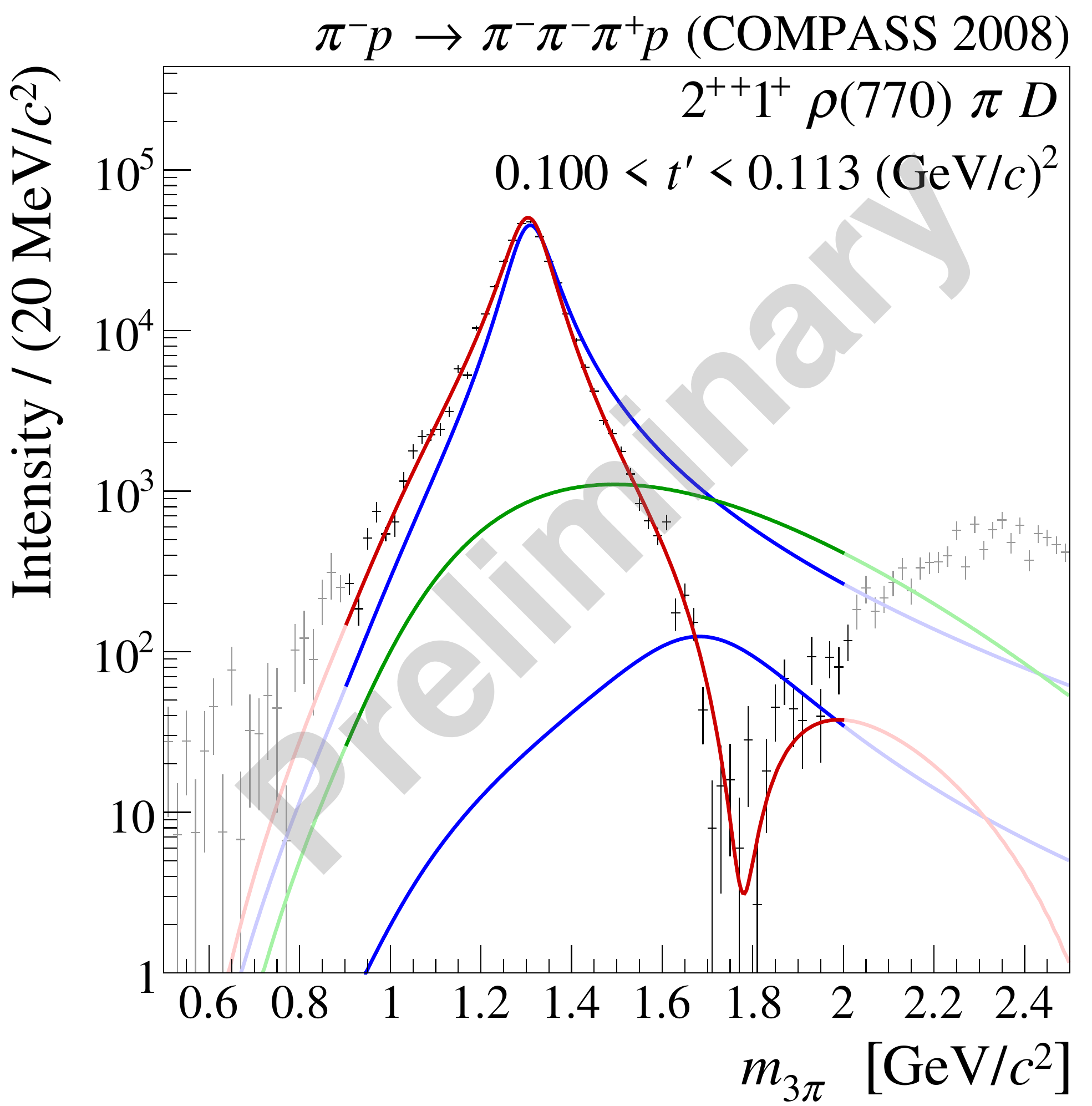}%
		\label{fit:res:2pp:rholowt}%
	}%
	\subfloat[]{%
		\includegraphics[width=\threePlotWidth]{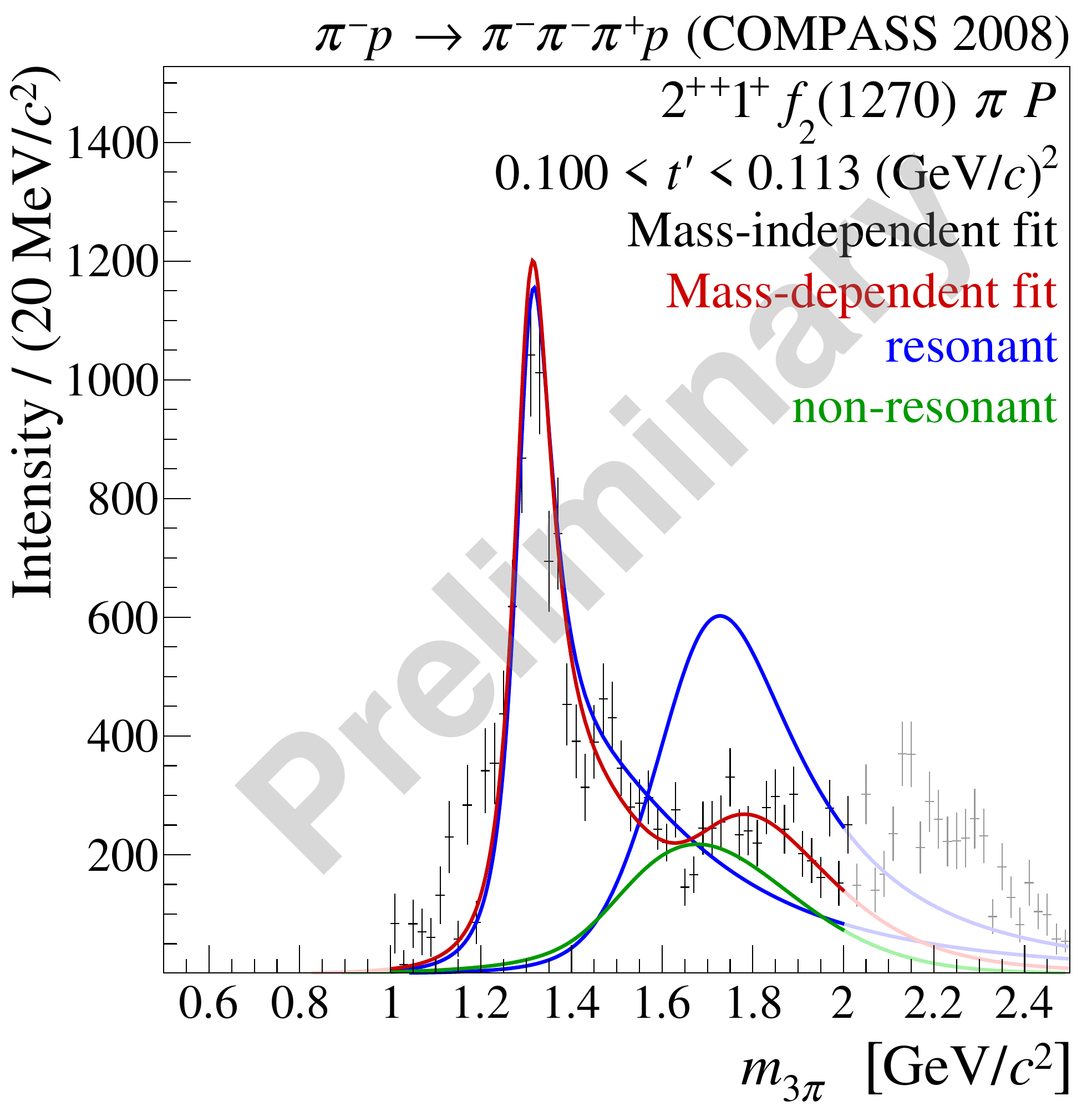}%
		\label{fit:res:2pp:f2lowt}%
	}%
	\subfloat[]{%
		\includegraphics[width=\threePlotWidth]{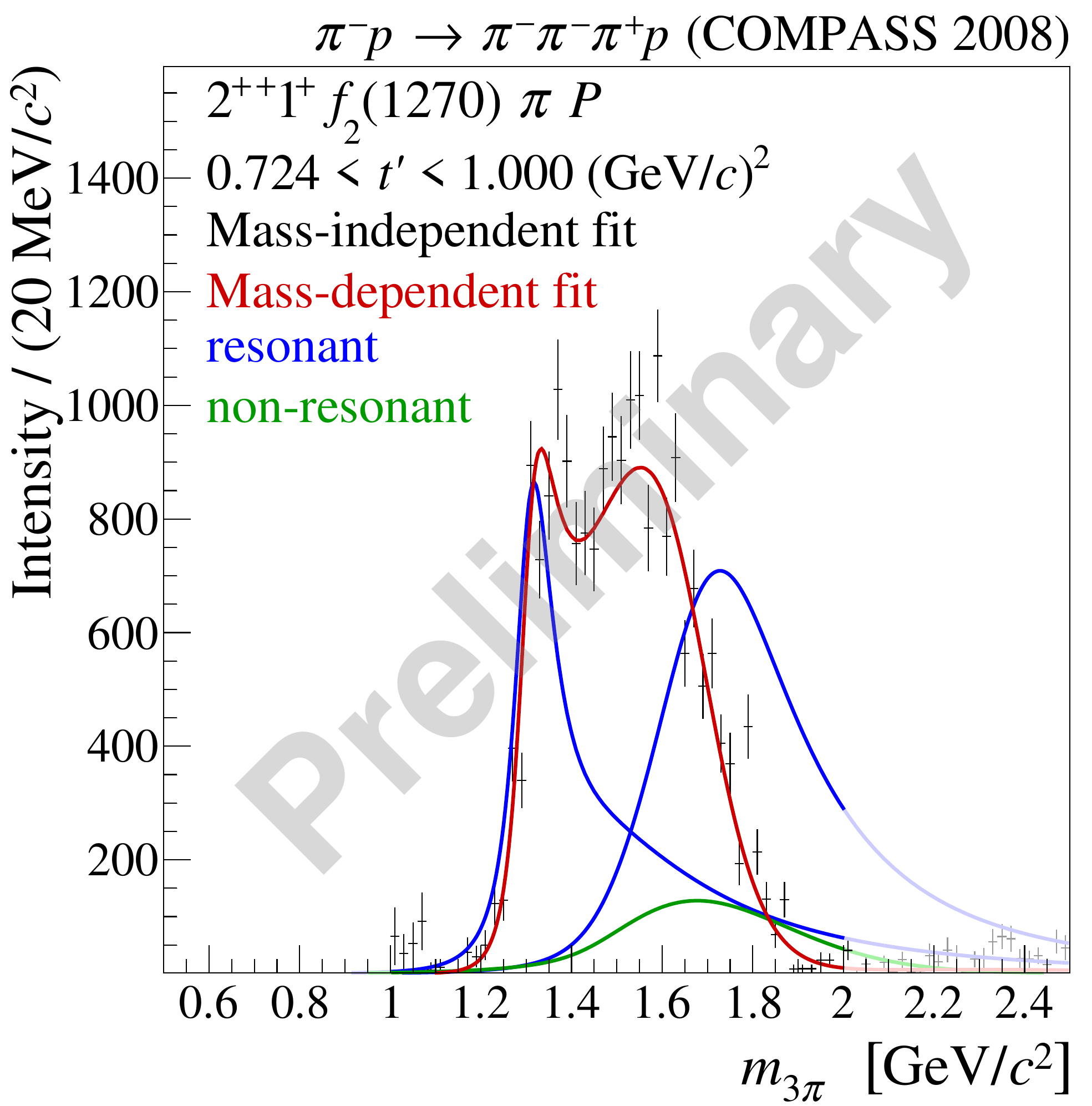}%
		\label{fit:res:2pp:f2hight}%
	}%
	\caption{(a) intensity distributions of the \Wave 2++1+\Prho\Ppi D wave in the lowest \tpr bin in log-scale. (b) and (c) intensity distributions of the \Wave 2++1+ \PfTwo \Ppi P wave in the lowest and highest \tpr bin, respectively. The data points show the result of the partial-wave decomposition. Errors are statistical only. The red curves represent the resonance model. The blue curves represent the resonances. The green curves represent the non-resonant components. The extrapolations beyond the \mThreePi fit range are shown in lighter colors.}
	\label{fit:res:2pp}
\end{figure}
The \Wave 2++1+\Prho\Ppi D wave exhibits a clear narrow peak in the mass region of the well-known \PaTwo resonance (see \Cref{fit:res:2pp:rholowt}). As it is expected, this peak is mainly described by the \PaTwo component with only small contributions from other components. Therefore, our estimate for the \PaTwo mass of $1314.2\,^{+1.0}_{-3.1}\,\si{\MeVcc}$ and width of $106.7\,^{+3.5}_{-2.4}\,\si{\MeVcc}$ is stable with respect to systematic effects and in agreement with previous measurements.

In addition to the \PaTwo peak, the \Wave 2++1+\Prho\Ppi D wave shows some evidence for a potential \PaTwoPr excited state. In the low \tpr-region, the observed dip at \SI{1.8}{\GeVcc} is reproduced well as a destructive interference between the \PaTwoPr component and the other components in this wave.
The \Wave 2++1+\PfTwo\Ppi P wave shows the strongest evidence for the \PaTwoPr. The shape of its intensity distribution is modulated strongly with \tpr (see \Cref{fit:res:2pp:f2lowt,fit:res:2pp:f2hight}), which is reproduced well by the model as a change of the interference pattern between the \PaTwoPr and the other components in this wave.
The measured \PaTwoPr mass and width parameters of $1674\,^{+140}_{-32}\,\si{\MeVcc}$ and $435\,^{+50}_{-15}\,\si\MeVcc$ show larger systematic uncertainties, as the \PaTwoPr is a small signal compared to the dominant \PaTwo. The PDG~\cite{Olive2016} lists the \PaTwoPr as ``omitted from summary table''~\cite{Olive2016}. The world average for its mass is in agreement with our observation, but our estimate for the width is \SI{240}{\MeVcc} larger than the world average.
A recent analysis of COMPASS data on the $\eta\pi$ final state using an analytic model based on the principles of the relativistic $S$-matrix also obtained a smaller width for the \PaTwoPr of $280 \pm 10(\text{stat.})\pm 70(\text{sys.})\,\si{\MeVcc}$~\cite{Jackura2017}\footnote{Their pole parameters for the \PaTwo are in good agreement with our results.}. However, the authors of Ref.~\cite{Jackura2017} only fitted the intensity spectra. In contrast to our analysis, the relative phase information and the evolution with \tpr were not taken into account.
\subsection{$\mathbf{\JPC = \onePP}$ resonances}
\begin{figure}[tbp]
	\subfloat[]{%
		\includegraphics[width=\threePlotWidth]{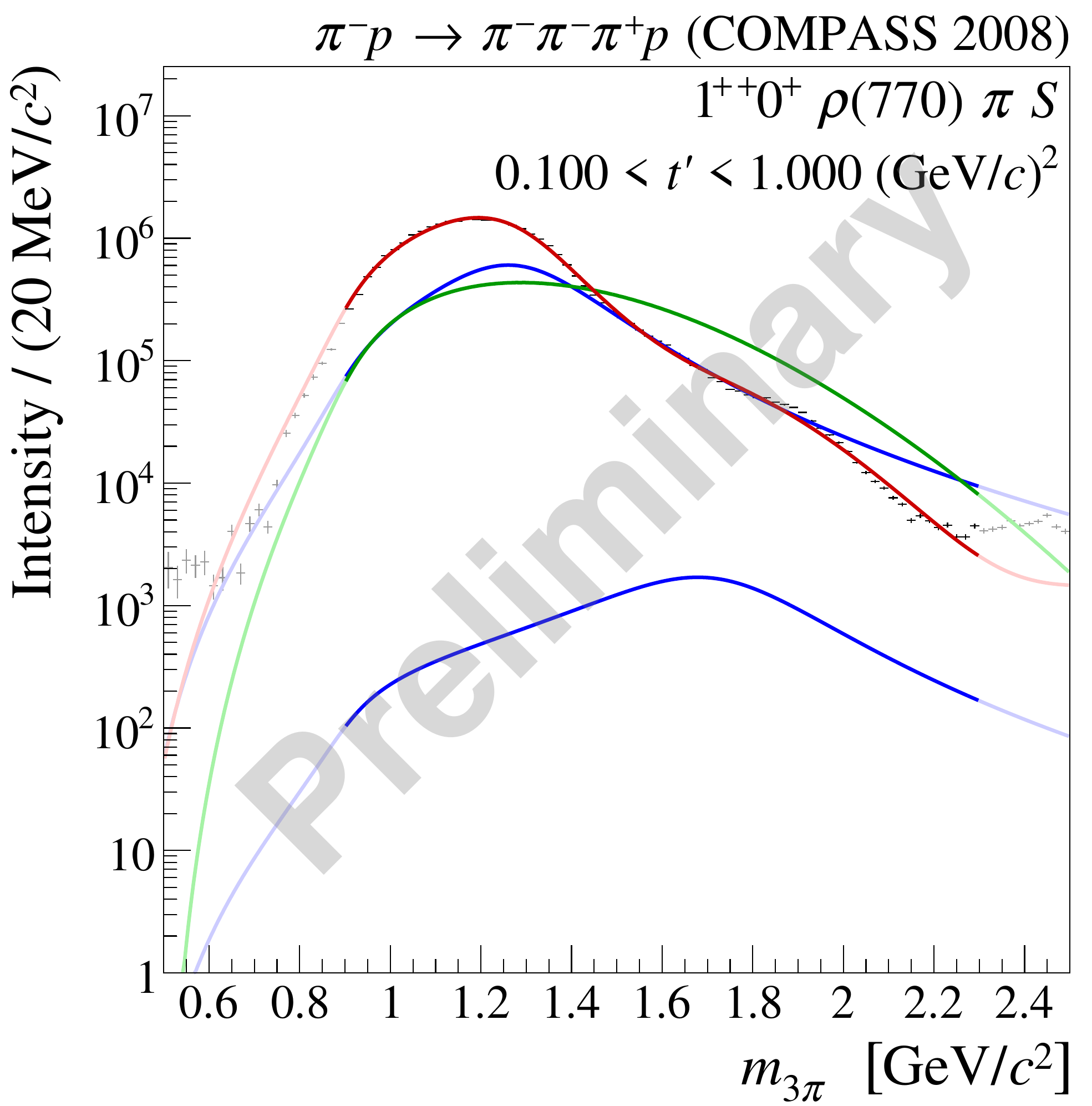}
		\label{fit:res:1pp:rho}%
	}%
	\subfloat[]{%
		\includegraphics[width=\threePlotWidth]{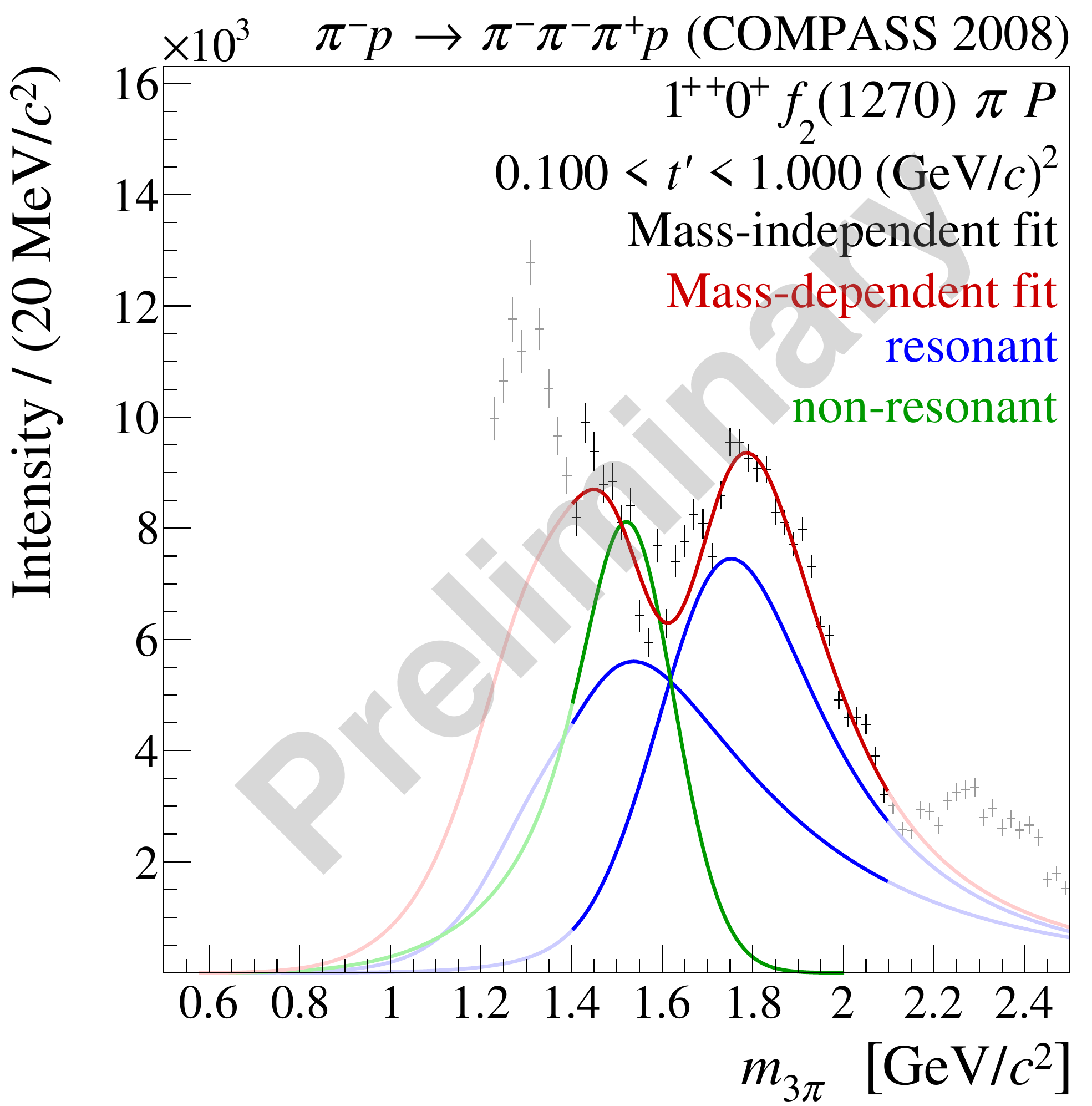}%
		\label{fit:res:1pp:f2}%
	}
	\subfloat[]{%
		\includegraphics[width=\threePlotWidth]{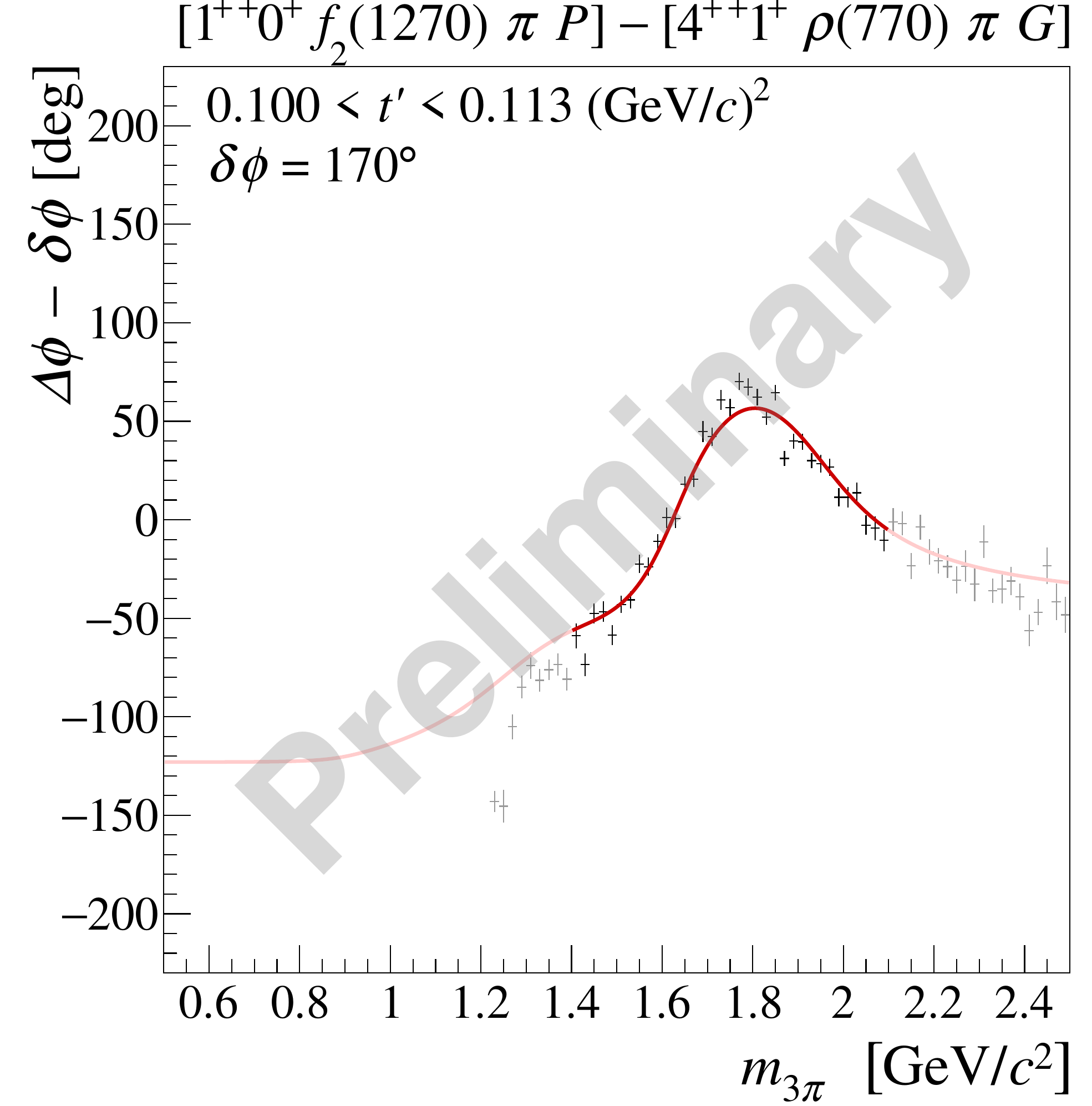}%
		\label{fit:res:1pp:f2phase}%
	}%
	\caption{\tpr-summed intensity distributions of the (a) \Wave 1++0+\Prho\Ppi S and the (b) \Wave 1++0+ \PfTwo \Ppi P waves. (c) shows the relative phase of the \Wave 1++0+ \PfTwo \Ppi P wave w.r.t. the \Wave 4++1+ \Prho\Ppi G wave. Same color code as in \Cref{fit:res:2pp} is used.}
	\label{fit:res:1pp}
\end{figure}
The \Wave 1++0+ \Prho\Ppi S wave contributes about \SI{30}{\percent} to the total intensity and  is the dominant signal in our data. It exhibits a broad peak in the intensity distribution at about \SI{1.2}{\GeVcc} (see \Cref{fit:res:1pp:rho}). This broad peak is described by the \PaOne component and the non-resonant component of similar intensity.
The resonance model cannot describe the details of the intensity spectrum of the \Wave 1++0+ \Prho\Ppi S wave within the small statistical uncertainties properly. This is mainly due to the large contribution of the non-resonant component, in combination with our lack of accurate knowledge about the non-resonant shape. This also leads to large systematic uncertainties of the measured \PaOne mass and width of $1298\,^{+13}_{-22}\,\si{\MeVcc}$ and $400\,^{+0}_{-100}\,\si{\MeVcc}$, respectively, which are in agreement with previous measurements.

Similar to the \twoPP waves, we observe the evidence for a potential \PaOnePr in the high-mass tail of the \PaOne, visible as a shoulder at about \SI{1.8}{\GeVcc} in the intensity spectrum of the \Wave 1++0+\Prho\Ppi S wave. The strongest evidence for the \PaOnePr is observed in the \Wave 1++0+ \PfTwo \Ppi P wave. It shows as a clear peak at about \SI{1.8}{\GeVcc} (see \Cref{fit:res:1pp:f2}) in combination with a rising phase motion in this mass region (see \Cref{fit:res:1pp:f2phase}). Both are reproduced well by the resonance model.
Our estimate for the \PaOnePr parameters are $m_0 = 1690\,^{+40}_{-70}\,\si{\MeVcc}$ and $\Gamma_0 = 534\,^{+124}_{-20}\,\si\MeVcc$. The PDG lists the \PaOnePr as ``omitted from summary table''. The world average for the mass is in agreement with our observation, but our estimate for the width is \SI{280}{\MeVcc} larger than the world average.

\subsection{$\mathbf{\JPC = 2^{-+}}$ resonances}
\begin{figure}[tbp]
	\subfloat[]{%
		\includegraphics[width=\twoPlotWidth]{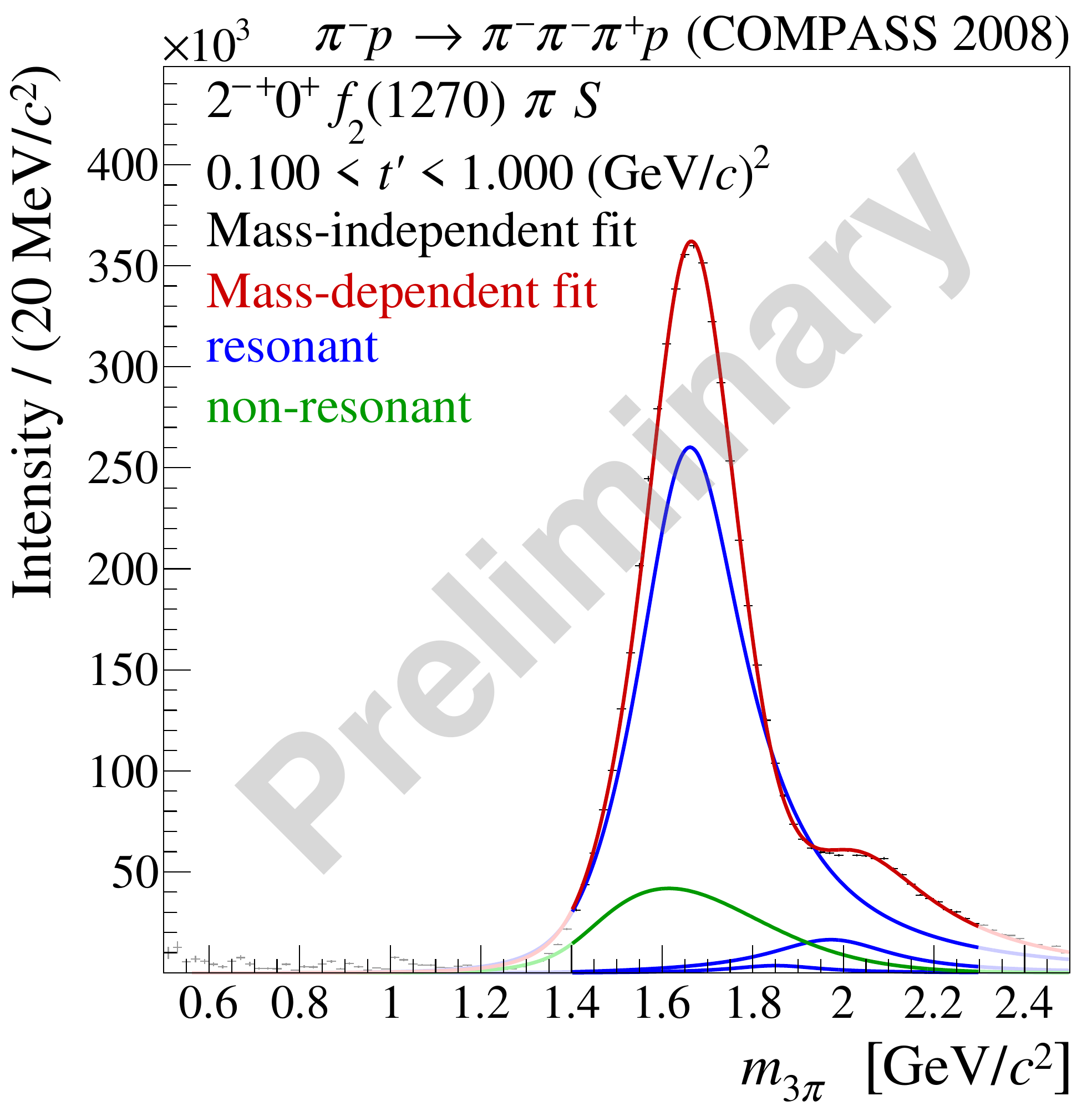}%
		\label{fit:res:2mp:f2S}%
	}%
	\subfloat[]{%
		\includegraphics[width=\twoPlotWidth]{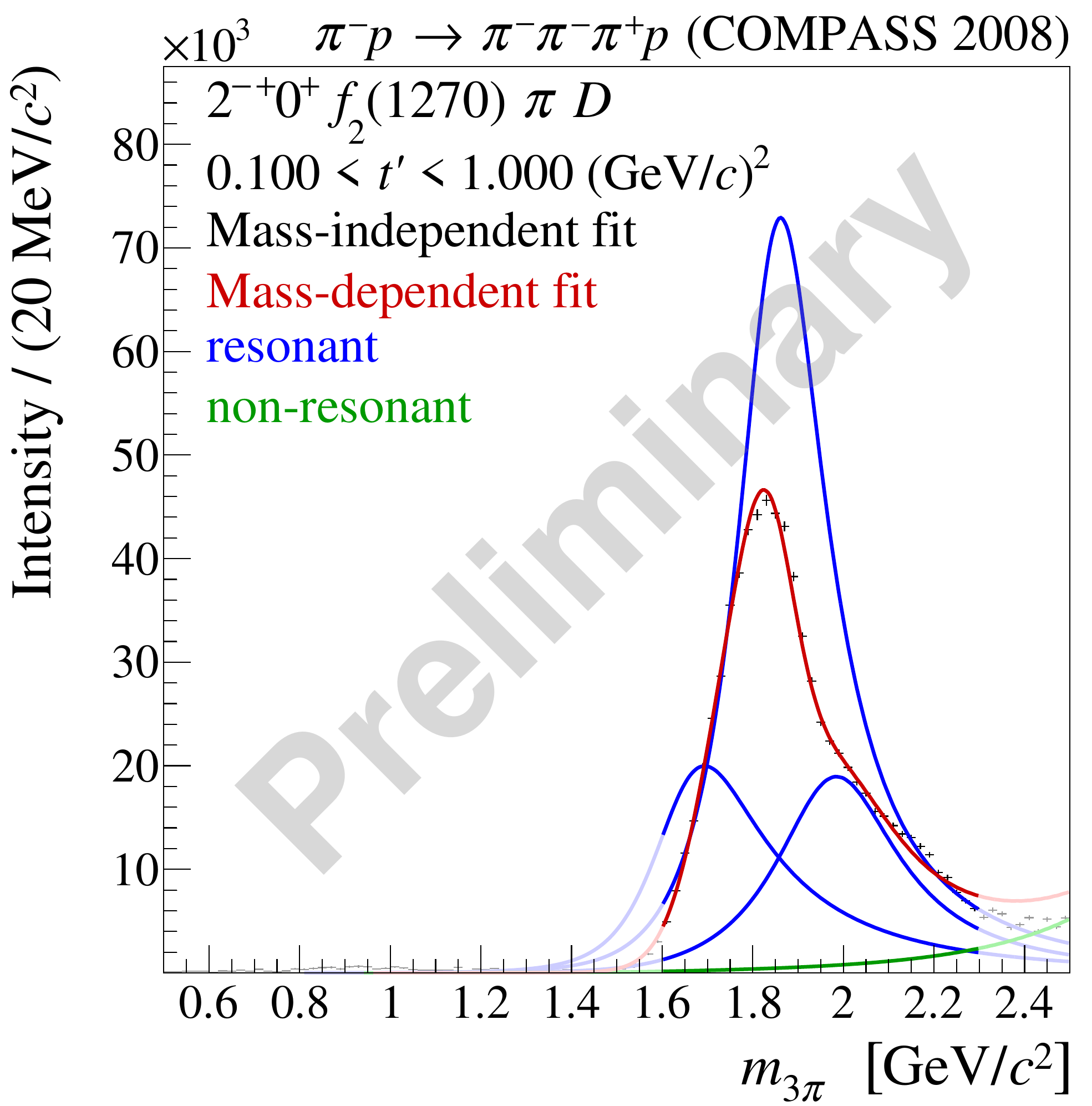}%
		\label{fit:res:2mp:f2D}%
	}\linebreak%
	\subfloat[]{%
		\includegraphics[width=\twoPlotWidth]{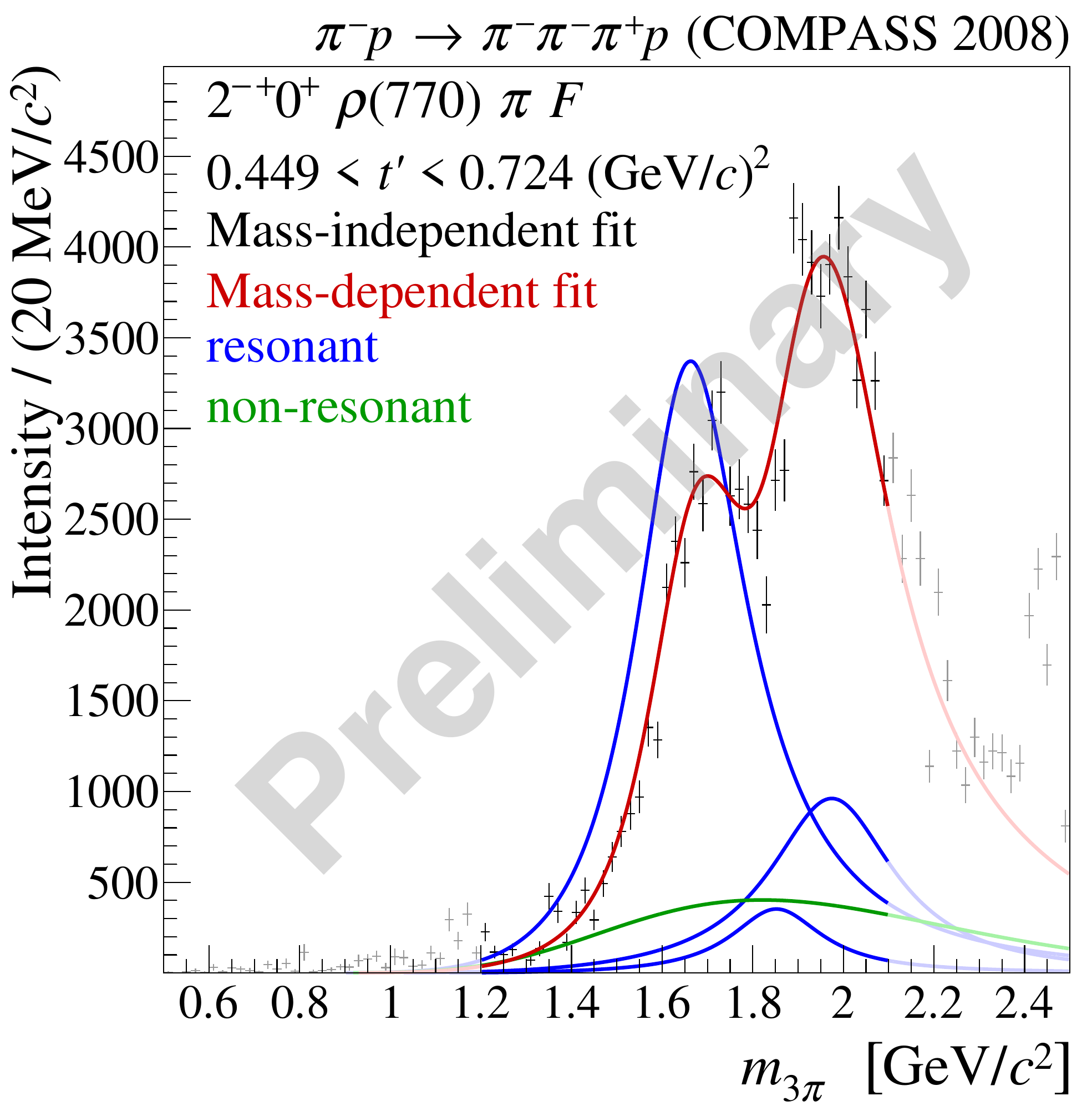}%
		\label{fit:res:2mp:rho}%
	}%
	\subfloat[]{
		\includegraphics[width=\twoPlotWidth]{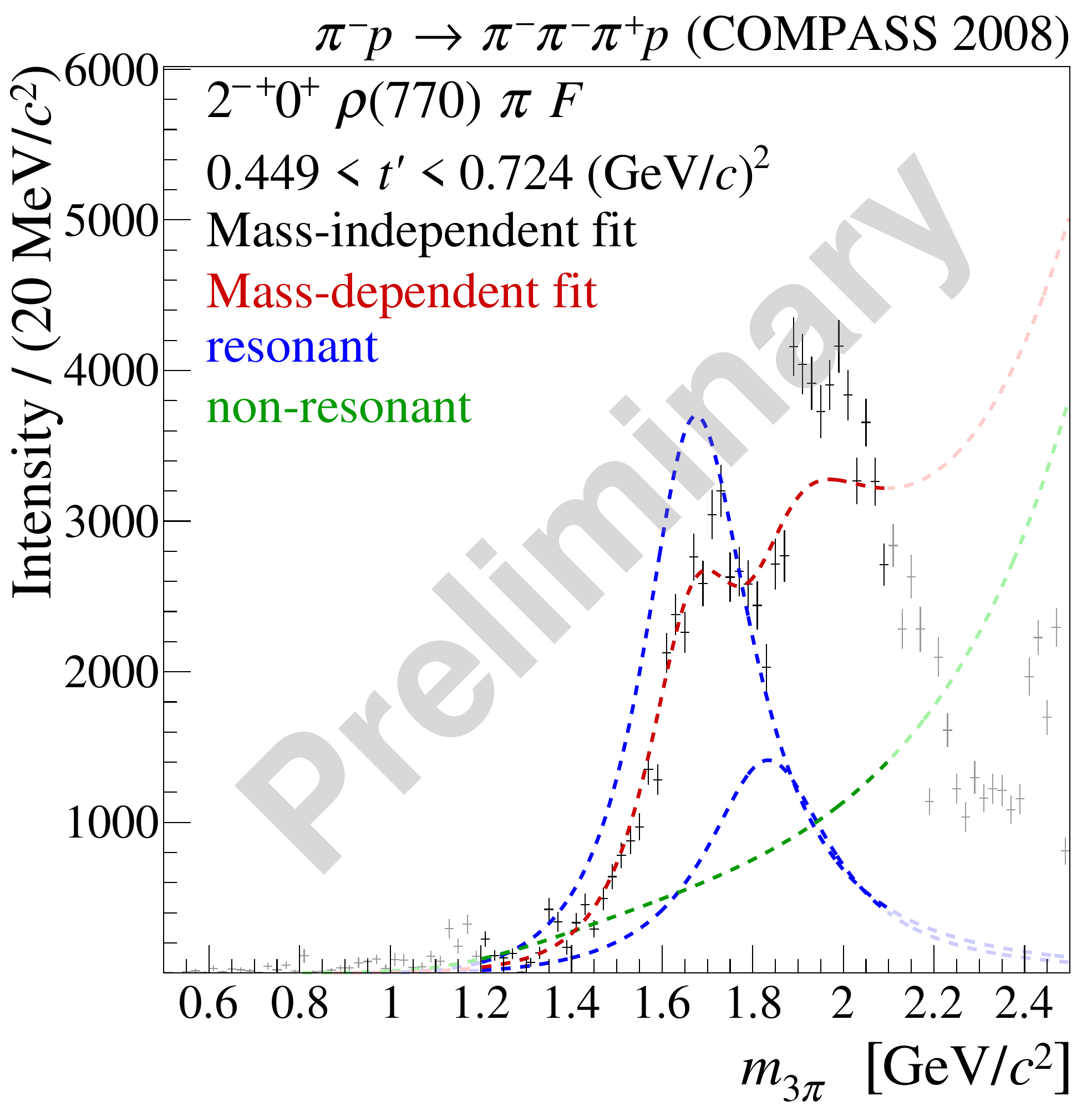}%
		\label{fit:res:2mp:rhostudy}%
	}%
	\caption{\tpr-summed intensity distributions of the (a) \Wave 2-+0+\PfTwo\Ppi S and (b) the \Wave 2-+0+\PfTwo\Ppi D waves. (c) and (d) show the \Wave 2-+0+\Prho\Ppi F partial-wave intensity distribution in the second highest \tpr bin. The curves in (c) show the standard fit, the dashed curves in (d) the result fo a fit without the \PpiTwoPrPr component. Same color code as in \Cref{fit:res:2pp} is used.}
	\label{fit:res:2mp}
\end{figure}
We include four partial waves with $\JPC=\twoMP$ into the resonance-model fit, each of which is parameterized by three resonance components and a non-resonant component.
The \Wave 2-+0+\PfTwo\Ppi S wave is the largest of the four. It exhibits a striking peak at about \SI{1.65}{\GeVcc} (see \Cref{fit:res:2mp:f2S}). This peak is reproduced well by a dominant contribution of the \PpiTwo component\footnote{We also include the \Wave 2-+{1}+\PfTwo\Ppi S wave with the spin-projection $M=1$, which has a similar peak as the $M=0$ wave.}.

The \Wave 2-+0+\PfTwo\Ppi D wave exhibits a striking peak as well. However, not at \SI{1.65}{\GeVcc}, but at about \SI{1.8}{\GeVcc} (see \Cref{fit:res:2mp:f2D}). The peak is described mainly by the \PpiTwoPr component. The low- and high-mass tails of the peak are described as an interference effect among the \PpiTwoPr, the \PpiTwo, or the \PpiTwoPrPr components.
 
The \Wave 2-+0+\Prho\Ppi F wave shows the strongest evidence for the \PpiTwoPrPr. We observe a double-peak structure (see \Cref{fit:res:2mp:rho}). The lower-lying peak is at about \SI{1.65}{\GeVcc} and is mainly described by the \PpiTwo component. The higher-lying peak is at about \SI{2}{\GeVcc} and is described by the \PpiTwoPrPr component. The relative strength between the peaks changes strongly with \tpr.
In order to study the significance of the \PpiTwoPrPr signal in our data, we performed a systematic study, in which we removed the \PpiTwoPrPr component from the resonance model, such that all four \twoMP waves are parameterized only by the \PpiTwo and \PpiTwoPr components and the non-resonant components. The result of this study is shown in \Cref{fit:res:2mp:rhostudy}. The \SI{2}{\GeVcc} mass region cannot be reproduced well without the \PpiTwoPrPr component\footnote{Additionally, the non-resonant term results in an unphysical shape in this study, because the fit uses the freedom of the non-resonant parameterization to compensate for the missing high-mass state.}. Furthermore, the \PpiTwoPr width increases in this study by about \SI{100}{\MeVcc}, and would be in contradiction to all previous observations.

\section{Discussion}
We have performed the so far largest resonance-model fit simultaneously and consistently describing  14 partial waves by a single model.
The huge amount of information condensed in this fit, in combination with the information from the \tpr-resolved analysis allows us to study ground  as well as excited states.
We observe signals of the \PaOnePr and \PaTwoPr, which are the first excitations of the \PaOne and \PaTwo,respectively. As both states appear in the high-mass tails of the dominant ground states, we extract their parameters with large systematic uncertainties. Our estimate for their widths is systematically larger in comparison to previous observations, which might indicate an imperfect separation of the wave components in the corresponding mass regions. We also observe, that the \Prho\Ppi decay mode of the \onePP and \twoPP waves is dominated by the ground states, while the excited states are more dominant in the \PfTwo\Ppi decay mode.
Also in the \twoMP waves we observe that the different resonances contribute with different relative strength to the different decay modes. Furthermore, we require a second excited state, the \PpiTwoPrPr, to describe properly all four \twoMP partial waves simultaneously.
To reduce the systematic uncertainties and further clarify the existence of higher states, models that are based on the principles of unitarity and analyticity are mandatory, especially in waves where the non-resonant contributions are important.

\printbibliography[heading=bibintoc]

\end{document}